**Causes and consequences of genetic background effects illuminated by integrative genomic analysis**

Christopher H. Chandler[*,§,1], Sudarshan Chari[*,§], David Tack[*,2], Ian Dworkin[*,§]

[*]Michigan State University, Department of Zoology, East Lansing, MI, 48824
[§]BEACON Center for the Study of Evolution in Action
[1]Present address: Department of Biological Sciences, SUNY Oswego, Oswego, NY 13126
[2]Present address: University of British Columbia, Centre for Plant Research, Vancouver, BC, Canada, V6T 1Z4

**Running title:** Genomic analysis of background effects

**Corresponding Authors:**

Dr. Ian Dworkin

Department of Zoology,

Michigan State University,

East Lansing, MI, 48824

Phone: 517 432 6733

Email: idworkin@msu.edu

Dr. Christopher Chandler

Department of Biological Sciences,

SUNY Oswego,

Oswego, NY

Phone: 315 312 2774

Email: christopher.chandler@oswego.edu

**Abstract**

The phenotypic consequences of individual mutations are modulated by the wild-type genetic background in which they occur. Although such background dependence is widely observed, we do not know whether general patterns across species and traits exist, nor about the mechanisms underlying it. We also lack knowledge on how mutations interact with genetic background to influence gene expression, and how this in turn mediates mutant phenotypes. Furthermore, how genetic background influences patterns of epistasis remains unclear. To investigate the genetic basis and genomic consequences of genetic background dependence of the *scalloped*[E3] allele on the *Drosophila melanogaster* wing, we generated multiple novel genome-level datasets from a mapping-by-introgression experiment and a tagged RNA gene expression dataset. In addition we used whole genome re-sequencing of the parental lines—two commonly used laboratory strains—to predict polymorphic transcription factor binding sites for SD. We integrated these data with previously published genomic datasets from expression microarrays and a modifier mutation screen. By searching for genes showing a congruent signal across multiple datasets, we were able to identify a robust set of candidate loci contributing to the background-dependent effects of mutations in *sd*. We also show that the majority of background-dependent modifiers previously reported are caused by higher-order epistasis, not quantitative non-complementation. These findings provide a useful foundation for more detailed investigations of genetic background dependence in this system, and this approach is likely to prove useful in exploring the genetic basis of other traits as well.

**Introduction**

Geneticists often strictly control their organisms' wild-type genetic backgrounds when experimentally dissecting genetic pathways. Although this tight control is necessary to avoid faulty inferences caused by confounding variables (e.g., BURNETT *et al.* 2011), it can often paint an incomplete or even incorrect picture; no genetic pathway or network exists in a vacuum. Instead, these networks occur in the context of all the alleles in the genome, which usually vary among individuals. There is substantial evidence that wild-type genetic background almost always modulates the phenotypic effects of mutations (e.g., MCKENZIE *et al.* 1982; THREADGILL *et al.* 1995; ATALLAH *et al.* 2004; MILLOZ *et al.* 2008; CHANDLER 2010; DOWELL *et al.* 2010; GERKE *et al.* 2010). The influence of wild-type genetic backgrounds also extends to interactions among mutations (REMOLD and LENSKI 2004; DWORKIN *et al.* 2009; WANG *et al.* 2013b), altering patterns of epistasis, and these complex interactions are likely widespread (CHARI and DWORKIN 2013). Alleles that influence many mutant phenotypes segregate in most natural populations, representing a potential source of cryptic genetic variation (POLACZYK *et al.* 1998; FÉLIX 2007; VAISTIJ *et al.* 2013). In many cases, this cryptic variation has been described phenomonologically, or via the partitioning of genetic variance components (GIBSON *et al.* 1999; DWORKIN *et al.* 2003; MCGUIGAN *et al.* 2011). However, its genetic basis remains poorly understood (DWORKIN *et al.* 2003; DUVEAU and FÉLIX 2012).

If our aim is to understand how a specific perturbation to a genetic network (e.g., a particular mutation) influences the phenotype outside a laboratory setting—a goal shared by multiple disciplines, from the genetics of disease to evolutionary biology— then we must go beyond simply characterizing the phenotypic consequences of background-dependent effects. That is, we need to understand both the causes and consequences of this genetic background dependence of mutational effects (CHANDLER *et al.* 2013). For instance, one study showed that specific quantitative trait nucleotide (QTN) alleles in naturally occurring yeast (*Saccharomyces cerevisiae*) isolates have complex phenotypic effects depending on both genetic background and environmental context (GERKE *et al.* 2010). However, to make predictions about how a particular QTN may influence trait variation in a novel genetic (or environmental) context requires a more mechanistic understanding of how the QTN allele interacts with the genetic background to influence phenotypes, whether at the level of gene expression, cellular changes, morphology, behavior, etc. (XU *et al.* 2013). This goal may involve mapping the background modifier loci to identify the genes and alleles responsible for the background dependence (e.g., CHANDLER 2010; DUVEAU and FÉLIX 2012), and testing which genes are mis-expressed (and how severely) as a result of the mutation in different genetic backgrounds (e.g., DWORKIN *et al.* 2009). Only with such data will we be able to identify common patterns across traits, taxa, and mutations.

Accomplishing these goals will require tractable models. Such models should provide the ability to examine specific mutations across multiple wild-type backgrounds and genomic resources to facilitate the analysis of the modifiers. One such system is the fly (*Drosophila melanogaster*) wing. Several mutations affecting wing development have background-dependent phenotypic effects (ALTENBURG and MULLER 1920; NAKASHIMA-TANAKA 1967; SILBER 1980; DWORKIN and GIBSON 2006; DEBAT *et al.* 2009), with selectable variation present in natural populations (THOMPSON 1975; CAVICCHI *et al.* 1989). In one dramatic example, the *scalloped*$^{E3}$ (*sd*$^{E3}$) mutation causes a moderately reduced, blade-like wing in the background of one common lab wild-type strain, Samarkand (hereafter, SAM), but a much more severely diminished wing in another common background, Oregon-R (ORE) (Figure 1; DWORKIN *et al.* 2009). These two parental strains are commonly used wild-types for *Drosophila* research, one primarily in molecular biology (ORE), and the other (SAM) in quantitative genetics, aging, and studies of recombination (HOFMANOVA 1975; HOFMANN *et al.* 1987; LYMAN *et al.* 1996; LYMAN and MACKAY 1998; LEIPS and MACKAY 2002; MACKENZIE *et al.* 2011). More recently it was used as the common line for an advanced intercross design for the *Drosophila* synthetic population resource (KING *et al.* 2012). Thus, identifying sequence polymorphisms and differences in gene expression between these two strains will facilitate studies of background dependence in other *Drosophila* traits.

The sets of genes whose expression is altered in wing imaginal discs by the *sd*$^{E3}$

mutation in each genetic background are largely non-overlapping, and partially independent from those genes that are differentially expressed between the two genetic backgrounds in a wild-type context (DWORKIN *et al.* 2009). The mechanisms underlying this background dependence, however, are poorly understood. Nevertheless, *sd* itself is well studied, making it a good system for investigating the mechanisms underlying background-dependent expression. SD encodes a TEA family transcription factor (TF) that forms heterodimers with multiple TF partners to regulate at least three biological processes (CAMPBELL *et al.* 1992; GUSS *et al.* 2013). Its best known role is in wing development, where SD interacts with the Vestigial (VG) protein (HALDER *et al.* 1998; PAUMARD-RIGAL *et al.* 1998; SIMMONDS *et al.* 1998; BRAY 1999; VARADARAJAN and VIJAYRAGHAVAN 1999; GUSS *et al.* 2001; HALDER and CARROLL 2001). In particular SD-VG regulates the expression of a number of genes influencing specification of cell fates as well as overall tissue growth in the wing imaginal disc (GUSS *et al.* 2001; HALDER and CARROLL 2001). Recently SD has been shown to be an important transcriptional co-factor with Yorkie (YKI), mediating hippo signaling to regulate growth and polarity of tissues (ZHAO *et al.* 2008; GOULEV *et al.* 2008; WU *et al.* 2008; ZHANG *et al.* 2008; 2009; REN *et al.* 2010; DOGGETT *et al.* 2011; NICOLAY *et al.* 2011; ZHANG *et al.* 2011; POON *et al.* 2012; CAGLIERO *et al.* 2013; KOONTZ *et al.* 2013; SIDOR *et al.* 2013). Finally, SD plays a role in the development of odorant and gustatory neurons in the adult (CAMPBELL *et al.* 1992; SHYAMALA and CHOPRA 1999; RAY *et al.* 2008), although the details of how it acts as a transcriptional regulator are less well known in this context.

Like other quantitative traits, the penetrance and expressivity of a mutation probably have a complex genetic basis. Indeed, in one high-throughput screen for conditionally essential genes in yeast, four or more modifier loci were necessary to explain the background dependence of numerous conditionally lethal gene deletions (DOWELL *et al.* 2010). These background polymorphisms may also have multifactorial consequences, such as pleiotropic effects on other unrelated traits (DUVEAU and FÉLIX 2012). In the case of our *Drosophila* wing model system, preliminary mapping efforts suggested the presence of a major-effect modifier on chromosome arm 2R near (but not in) *vg*. Yet the genetic background polymorphisms responsible for this background dependence remain unknown (DWORKIN *et al.* 2009). Unfortunately, due to the typically complex genetic basis of background dependence, fine mapping these modifier loci to specific polymorphisms, or even genes, remains challenging (ZHANG *et al.* 2013). However, we may be able to gain additional insight by integrating data from a variety of approaches. For instance, gene expression studies can identify genes whose expression is affected by a focal mutation in a background-dependent manner (DWORKIN *et al.* 2009; 2011), providing a complementary set of candidate loci. Then, using whole-genome re-sequencing to search for polymorphic transcription factor binding sites (TFBS) could help explain these background-dependent effects on gene expression.

   The primary goal of this current study was to identify the genomic regions that contribute to the genetic background effects of $sd^{E3}$. To do so, we integrated data from a variety of experimental sources to identify a robust set of candidate genes mediating the background dependence of the $sd^{E3}$ mutation. These sources included: (i) replicated introgression lines to map the loci contributing to background dependence; (ii) whole-genome re-sequencing of the introgression and parental lines to predict polymorphic SD transcription factor binding sites; (iii) expression profiling from a digital gene expression experiment and two existing microarray experiments; (iv) the results of a previously published screen for dominant modifiers of the $sd^{E3}$ phenotype. Despite earlier work suggesting that there may be a single modifier of large effect (DWORKIN *et al.* 2009), our mapping results indicate that this background dependence has a complex basis, encompassing several genomic regions each containing multiple candidate genes. Nevertheless, the additional data help narrow the list of candidates. In particular, patterns of allele-specific expression and SD binding site predictions suggest that cis-regulatory variants are not a major contributor to this background dependence. Finally, we extend recent results demonstrating that the majority of genetic interactions with $sd^{E3}$ are background dependent (CHARI and DWORKIN 2013), to determine whether these are the result of quantitative non-complementation or higher-order genetic interactions. Our results demonstrate that the majority of tested background-dependent modifiers cannot be explained simply by variation at the locus interacting with *scalloped*, ruling out simple quantitative non-complementation. Overall, our results lay the foundation for more detailed investigations into the mechanisms underlying genetic background dependence, and suggest that this approach should be useful for dissecting other complex phenotypes.

## Materials & Methods

**Mutations and wild-type strains:** As described in detail in DWORKIN *et al.* (2009), we introgressed the $sd^{E3}$ allele into the SAM and ORE wild-type strains via backcrossing for over 20 generations (each marked with $w^-$ to facilitate the introgression). Each wild-type strain (both with and without the $sd^{E3}$ mutation) was tested (using 8-10 molecular markers across the genome) to rule out contamination. Results from whole genome re-sequencing (see below) were consistent with these initial results.

**Generating short- and long-wing backcross lines for mapping:** To map the loci underlying the background dependence of the $sd^{E3}$ mutant phenotype, we used a phenotypic selection-based backcross procedure to introgress genomic regions that harbor the short-wing alleles into an otherwise long-wing genome (Figure 2). We generated a total of five independent backcross lines by crossing ORE $w$ $sd^{E3}$ flies to SAM $w$ $sd^{E3}$ flies, and crossing the F1s together. From the F2s, we selected flies with both long-wing (SAM-like) and short-wing (ORE-like) phenotypes. These flies were backcrossed to ORE $w$ $sd^{E3}$ and SAM $w$ $sd^{E3}$ flies, respectively, and the process was repeated. 12 cycles of backcrossing (24 generations) were completed for the short-wing backcross lines, and 20 cycles (40 generations) for the long-wing lines, with twenty flies of each parent strain used for backcrossing each generation (CHARI and DWORKIN 2013). We generated four independent short-wing replicates. Only one long-wing replicate remained at the final generation, as maintaining the long-wing phenotype while backcrossing to ORE $w$ $sd^{E3}$ was problematic because inheritance patterns broke down over successive generations: specifically, the long-wing phenotype ceased being dominant (ID, unpublished data).

**Phenotyping of wing size:** To compare wing lengths in the backcross lines, along with the "pure" SAM and ORE background $sd^{E3}$ strains, we grew each of the lines at 24°C, 65% RH on a 12:12 light cycle in a Percival incubator (I41VLC8) using our standard lab media. Once flies eclosed, they were stored in 70% ethanol. 20 wings were dissected from each genotype (10/sex) and mounted in 70% glycerol in PBS. Digital images were captured using an Olympus DP30BW camera mounted on an Olympus BW51 microscope at 40X magnification. Wing area was then estimated using a custom ImageJ macro that segments the wing blade from the rest of the image.

**Re-sequencing Oregon-R, Samarkand, and backcross lines:** We prepared genomic DNA from frozen whole flies using a Zymo Research Tissue & Insect DNA Kit (Irvine, CA) following the manufacturer's instructions and submitted samples to the Michigan State University Research Technology Support Facility for analysis on an Illumina Genome Analyzer II. SAM and ORE were each run on separate lanes while the five backcross lines were bar-coded, pooled, and run on a single lane. We obtained approximately 30,000,000 paired-end 75-bp reads (with an estimated insert size of 360 bp) for each lane, yielding 24-29x coverage for SAM and ORE and 5-7x coverage for each of the backcross lines.

We mapped reads to the *Drosophila melanogaster* reference genome (release 5.41) using bwa v. 0.6.0-r85 (Lɪ and Dᴜʀʙɪɴ 2009), allowing up to four mismatches (5%) per read. We used samtools v. 0.1.18 (Lɪ *et al.* 2009) to call SNPs in all samples simultaneously, and to generate consensus sequences for ORE and SAM. Sites with coverage higher than 100x were excluded from SNP calling to avoid false positives in repeat sequences. In downstream analyses of the backcross lines, we excluded indels and included only SNPs for which SAM and ORE were both scored as homozygous, but different from one another with high confidence (genotype quality scores 30 or greater as reported by samtools).

**Backcross analyses:** For each backcross line, we sought to determine whether the genome at a particular genomic location was inherited from SAM or ORE, so we focused our attention on SNPs at which SAM and ORE were scored as homozygous for different alleles. For each SNP, we estimated the frequency of the ORE (short-wing) and SAM (long-wing) alleles using custom Python scripts to count the number of sequence reads in each backcross line carrying each allele. We then plotted the proportion of reads displaying the ORE allele across the entire genome using a sliding window approach (with a window size of 10 kb and a step size of 2 kb). An alternate binning approach and various bin sizes yielded similar results.

**SD binding site analyses:** We sought to scan the ORE and SAM consensus genomes for genes that are predicted to be SD binding targets and in which sequence differences between SAM and ORE may alter putative SD binding sites. We first generated a position-weight matrix based on 23 known SD binding sites (Supplementary Tables 1 and 2) using the MEME suite (BAILEY *et al.* 2009). Next, for each annotated gene in the genome, we extracted its full sequence in ORE and SAM (including introns) plus 10kb of upstream and downstream flanking sequence. We used MotifScan.v6 (KIM *et al.* 2010) to generate a log-likelihood ratio (LLR) score for each gene, indicating the relative likelihood that the sequence contains one or more SD binding sites. We computed the difference between the LLR scores obtained from the SAM and ORE consensus sequences (ΔLLR). To identify genes with the strongest evidence of having predicted SD binding sites in SAM or ORE, or of being background-dependent SD targets, we derived empirical *p*-values for the LLR and ΔLLR scores by comparing each gene's LLR and ΔLLR scores to the 1000 other genes with the most similar GC content.

To complement these genome-wide binding site analyses, we also investigated the sequences of two known SD binding sites that were previously experimentally verified, in the genes *cut* and *salm* (GUSS *et al.* 2001; HALDER and CARROLL 2001), by extracting all sequence read data that mapped to those sites (using the mpileup command in samtools).

**Digital gene expression:** We used digital gene expression (MORRISSY *et al.* 2009) to identify genes that are mis-regulated in $sd^{E3}$ mutants and genes showing evidence of allelic imbalance in SAM/ORE "hybrids". This approach is similar to RNA-seq, but the cDNA is treated with a restriction enzyme (NlaIII) during library preparation, causing all sequence reads to begin at a restriction site. We first generated both mutant and wild-type F1 "hybrids" by crossing SAM $w$ $sd^{E3}$ flies to ORE $w$ $sd^{E3}$ flies. We dissected wing discs from wandering third instar larvae, and pooled ~200 wing discs to generate each sample. We generated two biological replicates each of both wild-type and mutant hybrid flies. We extracted total RNA using an Ambion MagMax-96 kit (Life Technologies) following the manufacturer's instructions. Further processing was performed at the Michigan State University Research Technology Support Facility, where samples were sequenced on an Illumina Genome Analyzer. After quality filtering, we obtained 7.6 to 8.1 million single-end, 17-bp reads from each sample.

We appended the NlaIII restriction site recognition sequence to the start of each sequence read and mapped sequence tags (unique reads) to the consensus ORE and SAM transcriptome sequences using bwa, allowing up to two mismatches (11%) per sequence. Sequences that mapped equally well to multiple genes were discarded from further analyses. All sequences that mapped to a single coding gene were included in analyses comparing wild-type flies to mutant flies. All sequence tags mapping to a single coding gene specifically in SAM or ORE were used in analyses of allelic imbalance. In cases in which either the SAM or ORE allele was missing from the sequence reads (i.e., only one of the two alleles was transcribed), we examined the SAM and ORE genome sequences to check for mutations in the NlaIII restriction sequence, because such a mutation would cause the mutation-bearing allele to be absent from our dataset even if it was being expressed. These cases were excluded from analyses of allelic imbalance.

To identify genes that are differentially regulated between wild-type and $sd^{E3}$ mutant flies, we used the edgeR Bioconductor package for R v. 3.0.8 (ROBINSON *et al.* 2010), using the exactTest() function, after estimating tagwise dispersions; all mapped reads were included in this analysis. We then computed q-values using the qvalue library (STOREY and TIBSHIRANI 2003). To test for allelic imbalance in genes in which sequence tags could be confidently assigned to either SAM or ORE, we compared binomial models using likelihood ratio tests. The null model assumed an equal probability (0.5) of a sequence read being derived from either SAM or ORE. The first alternative model allowed that probability to differ from 0.5, but was constrained to be the same in both wild-type and mutant flies. To identify genes in which the $sd^{E3}$ mutation alters the level of allelic imbalance, we also tested a second alternative model, in which the probability that a given sequence read was derived from SAM or ORE differed between wild-type and mutant flies.

**Microarrays:** We compared the results of our DGE experiment to two similar microarray experiments: one using Drosophila Genome Research Center arrays (DWORKIN *et al.* 2009), and one using a custom Illumina array (DWORKIN *et al.* 2011). Raw array data are available at NCBI GEO (GSE26706: GSM657471- GSM657487 & GSE13779: GSM346787- GSM346810). Briefly, genome-wide expression levels were quantified in wing discs of wandering third instar larvae with both wild-type and mutant ($sd^{E3}$) genotypes, and both the SAM and ORE backgrounds. For each gene, we asked whether expression is influenced by genotype (wild-type vs. $sd^{E3}$), genetic background (SAM vs. ORE), or an interaction between the two. For the DGRC dataset, we log-transformed and normalized the data and fitted mixed models using the sample genotype, genetic background, and their interaction as fixed effects, and array as a random effect, using the lmer function in R. For the Illumina dataset, we again log-transformed and normalized the data, and fitted linear models (using the lm function), with genotype, background, and their interaction as predictor variables.

**Modifier mapping:** We examined results from a prior study that screened two large autosomal deletion libraries (Exelixis and DrosDel) for dominant, background-dependent modifiers of the $sd^{E3}$ phenotype. Details are published elsewhere (CHARI and DWORKIN 2013). Briefly, we crossed males carrying each deletion to be tested to females carrying the $sd^{E3}$ allele in either the SAM or ORE background. We analyzed the wing phenotypes of the male offspring, which were hemizygous for $sd^{E3}$, heterozygous for the deletion of interest, and heterozygous for one of the two backgrounds (Supplementary Figure 4). As a control, we crossed SAM and ORE flies to the progenitor strains used to generate the deletion lines. We then scored wing phenotypes using a semi-quantitative scale (1-10; CHARI and DWORKIN 2013, TANAKA 1960), and used linear models to test for effects of each deletion, the genetic background, and their interaction on the wing phenotype. Deletions showing a significant interaction term were considered background-dependent modifiers. The raw data for the modifier mapping and for the subset of deletions used for the backcross-selection can be obtained from the Dryad Digital Repository (http://dx.doi.org/10.5061/dryad.4dt7c).

**Gene ontology analyses:** Using the sets of candidate genes from the expression, deletion modifier screen, binding site analysis, and mapping data, we tested for over-representation of gene ontology categories using WebGestalt (WANG *et al.* 2013a) with the enrichment test tool. We performed analyses for individual datasets (using the appropriate reference gene list), as well as a combined analysis in which candidate genes were identified by searching for genes showing a congruent signal across multiple datasets (with the full Ensembl *Drosophila* gene list as the reference). We used a Benjamini-Hochberg FDR, with a minimum of 3 genes per category. Given that for many genes we had only partial information (e.g., for some genes we had no or incomplete expression data, while for genes on the X chromosome, we had no modifier data) this list is incomplete. However, it still provides a useful view of the types of genes represented, and whether different subsets of biological processes are affected by the main genotypic effects of the $sd^{E3}$ allele and by its interactions with other loci. We also verified our results using the BioProfiling.de tool (ANTONOV 2011), which is a super-set of the analysis done by WebGestalt.

## Results

**Backcross mapping identifies genomic regions contributing to variation in the $sd^{E3}$ phenotype:** While the wings of the wild-type SAM and ORE flies are qualitatively similar, the effects of the $sd^{E3}$ mutation differ dramatically in each background (Figure 1). To identify regions of the genome associated with the long- and short-wing $sd^{E3}$ phenotypes, we backcrossed SAM or ORE $sd^{E3}$ flies to the alternate background while selecting for short- and long-wing phenotypes, respectively, each generation (Figure 2). Backcrossing was repeated for 12 cycles (24 generations) for the short- and 20 cycles (40 generations) for the long-wing backcrosses. This procedure is expected to introgress the ORE alleles that contribute to a short-wing phenotype into an otherwise SAM background, and vice versa. After backcrossing, we obtained one long-wing and four short-wing lineages phenotypically similar to $sd^{E3}$ in the "pure" SAM and ORE wild-type backgrounds (Figure 3), despite a predicted >95% replacement of the genetic background (Figure 4A).

To identify genomic regions harboring polymorphisms modulating genetic background effects, we used whole-genome re-sequencing of the parental and backcross strains to identify which genomic regions introgressed. While there were over 504,000 putative polymorphisms identified between the ORE, SAM, and the reference genome, we utilized only 92,006 SNPS that met our criteria (see Methods). We observed several regions that introgressed repeatedly in the short-wing backcross lines, with sizes ranging from just a few kb, up to ~5 Mb (Figure 4B, Supplementary Figure 1). A large portion of chromosome arm 3R introgressed, overlapping with several background-dependent modifier deletions of the $sd^{E3}$ phenotype. There were several additional regions, all of smaller sizes, on chromosome arm 3L that also consistently introgressed. Nearly 75% of the left half of chromosome arm 2R introgressed in some backcross lines, and approximately 5 Mb of chromosome 2L also introgressed across all short backcross lines, although each with clearly distinct recombination breakpoints. Several regions were less consistent. For instance, while a portion of chromosome arm 2L introgressed repeatedly, the breakpoints around the introgressed region were variable among lines (Figure 4B, Supplementary Figure 1), with one replicate line exhibiting nearly complete introgression of the entire chromosome arm.

In the long-wing backcross line, we expected the opposite genomic pattern from the short-wing lines, i.e., mostly ORE alleles, except where long-wing SAM alleles had introgressed. We identified only a few regions exhibiting this pattern (Figure 4, Supplementary Figure 1), some of which corresponded to the introgression regions for the short-wing backcross lines: one on the left half of chromosome arm 2L, one on the left half of 2R, one on 3L, and one on 3R. Surprisingly, in some regions where an ORE segment had introgressed into an otherwise SAM chromosome in the short-wing backcross lines, the same allelic distribution was observed in the long-wing backcross line, rather than the opposite pattern as expected for background modifiers of the $sd^{E3}$ phenotype, e.g., a small region near the right end of 3R (Figure 4, Supplementary Figure 1).

**Strain-specific polymorphisms in predicted SD binding sites weakly correlate to differences in gene expression:** To identify putative polymorphic SD transcription factor binding sites that may contribute to the background-dependent effects of the $sd^{E3}$ mutation on gene expression, we searched our re-sequenced SAM and ORE genomes with MotifScan v6 (Kɪᴍ *et al.* 2010) for predicted SD targets. Using the log-likelihood ratio (LLR) scores generated by MotifScan, we computed an empirical p-value for each gene by comparing its LLR score to the LLR scores of 1000 genes most similar in GC content (see **Methods** for details). 163 genes were predicted to be targets of regulation by SD (p ≤ 0.01 in either SAM or ORE; Table 1; see Dryad doi:10.5061/dryad.1375s for full results). Of these, 131 were common to both genetic backgrounds, 16 were SAM-specific, and 16 were ORE-specific. Genes predicted to be SD targets in at least one of the two backgrounds showed significant overlap with genes identified as being mis-expressed in $sd^{E3}$ flies in one of three gene expression datasets (the DGRC dataset; see **Integrating multiple datasets…** below); however, the degree of overlap was not especially strong (using a more relaxed significance threshold to increase the total strength of the signal in the data, observed overlap = 16 genes out of 10380 common to both datasets, of which 146 were differentially expressed in DGRC at q ≤ 0.05 and 535 were significant in the binding prediction analysis at p ≤ 0.05; p = 0.004 in a randomization test).

In addition, we also searched for genes that showed evidence of differential affinity for SD between the SAM and ORE genomes by identifying genes with exceptional differences in LLR scores computed using the SAM and ORE sequences, regardless of whether or not the gene was predicted to be a significant SD target in our first analysis. Using this method, we identified 149 genes predicted to show differential affinity for SD ($p \leq 0.01$), although only 37 of these were predicted to be significant SD binding targets ($p \leq 0.01$).

We also manually checked two experimentally verified SD binding sites (Guss *et al.* 2001; Halder and Carroll 2001) for polymorphisms. The genes regulated by both of these binding sites were predicted to be SD binding targets by our analyses (*cut*: p = 0.018 in SAM and p = 0.022 in ORE; salm: p =  0.026 in SAM and p = 0.023 in ORE), but neither was predicted to show differential affinity for SD binding. There was no evidence of substitutions in the SD binding site that regulates *cut* expression (Supplementary Table 3). However, ORE (but not SAM) showed evidence of three substitutions in the known SD binding site that regulates *salm* (Supplementary Table 4). *salm* was one gene that was verified (via in situ hybridization) to be differentially expressed between SAM $sd^{E3}$ and ORE $sd^{E3}$ (Dworkin *et al.* 2009).

**Allelic imbalance in gene expression is unlikely to contribute to the phenotypic effects of genetic background:** We generated digital gene expression (DGE) data—sequences from short mRNA tags (cleaved by a restriction enzyme)—in SAM/ORE "hybrid" flies with both wild-type (ORE $sd^+$/SAM $sd^+$) and $sd^{E3}$ (ORE $sd^{E3}$/SAM $sd^{E3}$) genotypes. We used this to identify (i) genes that are mis-regulated in the presence of the $sd^{E3}$ mutation; (ii) genes showing evidence of allelic imbalance, i.e., unequal transcription of the SAM and ORE alleles of the same gene; and (iii) genes showing evidence of genotype-dependent allelic imbalance.

A total of 59,057 mappable unique sequence tags representing 11,267 transcripts from 10,312 genes were identified. Of these, 2,078 sequence tags representing 1,639 genes displayed significantly different expression levels between wild-type and $sd^{E3}$ flies at a threshold of $q \leq 0.001$. This set of differentially expressed genes contained a disproportionate number of genes annotated as being involved in spindle organization and mitotic spindle organization, microtubule cytoskeleton organization, spindle elongation and mitotic spindle elongation, and gene expression, among others (Supplementary Figure 2, Supplementary Table 5).

Because DGE was performed in SAM/ORE "hybrid" flies, we also wanted to test for allelic imbalance, i.e., differential expression of the two copies of a gene within an individual. There were 2,583 mappable unique sequence tags representing 1,970 transcripts and 1,937 genes for which we could distinguish between the SAM and ORE alleles. Of those, 392 tags representing 378 genes showed evidence of allelic imbalance at a threshold of q ≤ 0.001 (Supplementary Table 6). Genes showing evidence of allelic imbalance were not enriched for any GO categories after adjustment for multiple comparisons.

We identified 29 genes showing evidence of a difference in the level of allelic imbalance between wild-type flies and $sd^{E3}$ mutant flies (q ≤ 0.001; Table 2). There was no more overlap than expected by chance between these genes and genes showing evidence of differences in SD binding affinity between SAM and ORE (p = 0.52 using a randomization test, again with a more liberal significance cutoff; observed overlap = 11 genes out of 1063 common to both datasets, of which 156 were significant in the DGE dataset at p ≤ 0.01 and 71 were significant in the binding prediction dataset at p ≤ 0.05). Similarly, these genes did not show any more overlap than expected by chance with the genes showing evidence of a genotype-by-background interaction influencing expression levels in either of two previously published microarray datasets (p = 0.11 and p = 0.93). Considering *sd* genotypes separately, 337 genes showed evidence of allelic imbalance in wild-type flies (at the same threshold), and 146 genes showed evidence of allelic imbalance in $sd^{E3}$ flies, with 26 of these genes overlapping (all at q ≤ 0.001). Again, GO terms showed no significant enrichment after correction for multiple comparisons. Lastly, there did not seem to be any overall, consistent bias towards the expression of one background's alleles, or any consistent change in allelic imbalance between wild-type and mutant flies (Figure 5, Supplementary Figure 3). Thus, despite some evidence of allelic imbalance, it does not appear to correlate well with other features of the background dependence. Whether this is a function of the underlying biology or the limited representation of the short sequence tags (limiting our ability to discern RNA from each parental allele) is unclear.

**The majority of background dependent genetic interactions with $sd^{E3}$ cannot be explained by quantitative non-complementation across the wild-type strains:** A previous deletion screen for dominant modifiers of the $sd^{E3}$ phenotype showed that the majority of observed modifiers display background dependence (CHARI and DWORKIN 2013). That is, most deletions that alter the $sd^{E3}$ phenotype do so differently when expressed in the SAM and ORE genetic backgrounds. However, that study was unable to distinguish between two competing hypotheses to explain this result. The first is quantitative non-complementation, i.e., second-order epistasis between the $sd^{E3}$ allele and the hemizygous background allele uncovered by the deletion (PALSSON and GIBSON 2000; MACKAY *et al.* 2005). Alternatively, background-dependent modifiers may be explained by third-order (or higher) epistasis between the $sd^{E3}$ allele, the deletion itself, and one or more genetic background alleles elsewhere in the genome. To evaluate these hypotheses, we used data from crosses between 31 deletions showing background-dependent effects to the short- and long-wing introgression lines (Figure 2, Supplementary Figure 4). If the modifier deletion's background dependence is due to quantitative non-complementation, then the wing phenotype will depend on the genetic background (SAM or ORE) the fly carries across from the deletion. On the other hand, if this background dependence is due to higher-order epistasis, the phenotype will also depend on alleles present at other locations in the genome, not just alleles uncovered by the deletion.

We can therefore test for quantitative non-complementation by comparing the genotypes and phenotypes of flies carrying a deletion of interest and the $sd^{E3}$ mutation in (i) a pure SAM or ORE parental background, versus (ii) an introgression background. Specifically, if flies from these two cases exhibit the same modification of the wing phenotype, but are hemizygous for different alleles at the deletion locus, then alleles elsewhere in the genome must be influencing how the deletion modifies $sd^{E3}$'s effects, and we can rule out a "simple" two-way interaction (Supplementary Figure 4). Similar logic applies for the situation in which the lineages have distinct phenotypes but the same allele opposite the deletion. For ~80% of the deletions tested, at least one cross between a backcross and the deletion provided evidence that second-order epistasis is insufficient to explain the background dependence of the modifier's effects (Table 3). Thus, the majority of modifier loci are background dependent because of third- or higher-order epistasis between the focal mutation, the modifier, and additional loci elsewhere in the genome.

**Integrating multiple datasets generates a robust set of candidate loci:** We also took advantage of several additional datasets. First, we used two previously published microarray datasets comparing gene expression profiles between wild-type and $sd^{E3}$ flies in both the SAM and ORE genetic backgrounds to identify genes whose expression levels are influenced by $sd$ genotype, genetic background, and their interaction (DWORKIN *et al.* 2009; 2011). Second, we examined the results of the screen for autosomal dominant modifiers of the $sd^{E3}$ mentioned earlier (CHARI and DWORKIN 2013).

To identify candidate genes for further study, we searched for genes displaying a common signal in multiple datasets: (i) those falling within a region showing evidence of introgression in the short-wing backcross lines; (ii) genes occurring within modifier deletions or background-dependent modifier deletions; (iii) differentially expressed genes in the microarray and DGE datasets (specifically, genes whose transcript levels were influenced by $sd$ genotype, a genotype-by-background interaction in the microarray datasets, or genotype-dependent allelic imbalance in the DGE dataset); (iv) genes predicted to show differential affinity for SD binding between SAM and ORE. Because significant but incongruent effects in independent expression datasets (e.g., up-regulation in $sd^{E3}$ flies in one dataset, and down-regulation in another dataset) could be a cause for concern, we also "flagged" genes showing evidence of such inconsistency. We then plotted all of these sources of data on a common set of axes (Figure 6, Supplementary Figure 5) and selected a set of robust candidates supported by multiple independent data types, including multiple expression datasets. At this stage, we used liberal cutoffs to identify "significant" genes (q < 0.05 for most datasets;

p < 0.05 for SD binding and the Illumina microarray dataset; and short-wing allele frequency ≥ 0.6 for the mapping-by-introgression dataset), because we were interested only in those genes showing a consistent signal across multiple datasets.

By integrating all these results, we generated a set of "high confidence" candidates (Table 4) that mediate or are modulated by either the effect of genotype ($sd^{E3}$ vs. wild-type) or an interaction between genotype and wild-type genetic background. Not surprisingly, some well-known genes that interact with or are regulated by SD, such as *vg*, *bi/Omb*, *fj*, *Dll*, and *chico*, are found among these candidates. More generally we observed over-representation of genes involved with *Drosophila* wing development (Supplementary Figure 6) for both the main and background-dependent interaction effects. There was also modest enrichment of microtubule associated genes for genotypic effects, and RAS GTPases and intracellular membrane-bounded organelle genes for the interaction effects, although it is currently unclear whether these distinctions reflect a true difference among the types of gene products that mediate background dependence.

## Discussion

The genomic context in which an allele finds itself can profoundly influence that allele's phenotypic consequences. Here, we have integrated genomic data from multiple experiments to examine how that context modulates the phenotypic effects of a specific mutation, the *scalloped*$^{E3}$ allele, on the *Drosophila melanogaster* wing. Although these datasets did not always yield a congruent signal, we were able to identify several strong candidate regions likely to be involved in mediating the background dependence of this allele's effects on wing development. Importantly, each dataset in isolation yielded a large set of candidates; only by examining these disparate sources together were we able to identify a robust and practical set of candidates for more focused study.

**Mapping by introgression:** Using introgression to map the polymorphisms responsible for modulating $sd^{E3}$'s consequences on wing phenotypes points to a complex genetic basis for this background dependence. However, pinning this background dependence on specific polymorphisms or even specific regions is complicated by several difficulties. First, large chromosomal blocks introgressed in many cases, each containing many genes. One possible explanation for this result is selection for multiple linked alleles, which would drag the entire segment to fixation, essentially building a long large-effect quantitative trait locus out of many smaller ones. In this case, additional crosses to generate recombination events within these regions may prove fruitful in pinpointing the polymorphisms responsible for $sd^{E3}$'s background-dependent effects.

An alternative, but not necessarily mutually exclusive, explanation is that the causal polymorphisms lie within polymorphic inversions, suppressing recombination within these blocks. A related and intriguing hypothesis is that rearrangements themselves may be responsible for the background dependence, making it not only impossible but also illogical to map the background modifiers to individual SNPs. By looking for sequence read pairs that mapped to discordant genomic positions, we found little evidence that inversions relative to the reference *D. melanogaster* genome are present in these regions, as each putative inversion is supported by only a few discordant read pairs, there are multiple conflicting inversions within each strain, and few putative inversions appear polymorphic between SAM and ORE (Supplementary Figure 7). We therefore think this hypothesis is unlikely to explain the large introgression blocks.

In a few instances, the same allele became fixed in both short- and long-winged backcross lines, even though alternate alleles were expected to be selected in these contrasting treatments. These loci may have been influenced by unintended selection for alleles influencing viability, rather than for wing phenotypes. Indeed, such unanticipated genetic effects have been observed in other studies (e.g., SEIDEL *et al.* 2008; ROSS *et al.* 2011; KING *et al.* 2012). Moreover, there was some inconsistency among introgression lines. We therefore propose that studies intending to map trait variation using an introgression-and-resequencing approach (e.g., EARLEY and JONES 2011) must include reciprocal crosses and multiple replicates. A single replicate of this introgression in only a single direction, for instance, would have missed some introgression regions, and some of the identified regions of introgression may have more to do with viability than with the trait of interest.

It is clear that multiple loci are involved, and problems generating multiple replicates of the long-winged backcross line (not shown) may indicate complex epistatic interactions between these loci. Although widely used with great success for Mendelian traits, the mapping-by-introgression approach adopted in the first part of our study may not be well suited to mapping trait variation with such a complex genetic basis. Nevertheless, in conjunction with the results of additional experiments such as gene expression data, this approach can still provide useful information that helps identify candidate genes for further investigation.

**Majority of background-dependent modifiers cannot be explained by quantitative non-complementation:** An earlier study showed that the majority of the $sd^{E3}$ allele's modifiers have effects that are background dependent (CHARI and DWORKIN 2013). Here, we extend that finding by providing evidence that these modifiers' background dependence is, in most cases, due to higher-order epistasis—interactions between the focal mutation (in this case, $sd^{E3}$), the modifier itself (in this case, a deletion), and alleles elsewhere in the genome. This finding is consistent with growing evidence that higher-order epistasis is prevalent (WEINREICH *et al.* 2013), and that what may initially seem to be a two-way interaction is often a more complex interaction involving additional loci or environmental influences (WHITLOCK and BOURGUET 2000; GERKE *et al.* 2010; WANG *et al.* 2013b; LALIĆ and ELENA 2013). To our knowledge this is the first attempt at combining such genetic and genomic data to infer the order of epistatic interactions (or at least to rule out lower-order interactions). While this approach can only be used for genetically and genomically tractable systems, it does allow for making a clear inference even in the absence of the exact identity of some of the interacting partners. Prior results are consistent with the hypothesis that the background loci altering interactions between $sd^{E3}$ and its modifiers are the same as those mediating the background dependence of $sd^{E3}$ itself (CHARI and DWORKIN 2013). Thus, identifying the latter should provide insights into the former, and further experiments can shed light on the precise nature of these interactions.

**Binding site analyses:** Using the whole-genome sequence data for SAM and ORE, we identified a large number of predicted SD binding sites. More relevant for this study, we also identified a moderate number of genes whose regulatory elements are predicted to bind SD differentially between the SAM and ORE genetic backgrounds. Nearly all of these await experimental validation. In addition, binding predictions were only weakly correlated with expression differences. This result should not be too surprising, as the set of genes whose expression changes in the presence of the $sd^{E3}$ mutation includes both direct and indirect targets of SD. Nevertheless, two validated targets of SD binding were picked by our approach as predicted targets of SD regulation. While one of these was manually found to contain three polymorphisms in the SD binding site, this differential SD binding affinity was missed by our analysis. Moreover, while exploring different binding site prediction methods, we also found that results were highly dependent upon the approach used as well as on the choice of specific position weight matrix (not shown).

Combined, these observations suggest that current binding site prediction approaches may not be specific or sensitive enough to produce completely reliable sets of candidate regulation targets on their own. Even so, they still seem to possess enough predictive power that, when combined with other sources of data, they can strengthen the evidence for the involvement of some genes in a biological process of interest. For instance, several of the candidate genes that were already strongly supported by other sources of data were also predicted to be SD binding targets or to possess polymorphic SD binding sites (Table 4).

**Gene expression:** Though the direction of effects was mostly consistent, the same genes were not always identified as "significant" in independent expression datasets. These differences may be explained by the unique technologies used with different biases, as well as distinct experimental designs. For example, the DGE experiment could only measure the abundance of transcripts containing the restriction sequence (CATG), may have been subject to biases in mapping reads to the reference genome, and our ability to detect background-dependent expression was contingent on the presence of SNPs in our sequence reads derived from SAM-ORE hybrid flies. The microarray datasets, on the other hand, used only flies with either a pure SAM or ORE background, and each array could only measure the expression of the specific transcripts complementary to the arrays' probes, which may also have been influenced by polymorphisms between the SAM and ORE sequences and the probe sequences.

**Allelic imbalance:** There was very little overlap between genes showing evidence of allelic imbalance in SAM/ORE hybrid flies, genes showing evidence of background-dependent expression levels in pure SAM and ORE flies, and genes containing polymorphic predicted *scalloped* binding sites. The lack of overlap between the first two sets of genes is not surprising, because allelic imbalance must be due to cis-regulatory variation, whereas expression differences between pure SAM and ORE flies may be due to either cis- or trans-regulatory differences. Moreover, our ability to detect allelic imbalance in the DGE dataset was limited by our ability to identify SNPs in our relatively short sequence tags, and by a lack of statistical power for low-abundance transcripts. The comparison between genes showing allelic imbalance and genes predicted to have polymorphic SD binding sites is further hindered by imperfect binding site predictions, and because TFBS can reside quite far from the gene they regulate. The lack of strong congruence between these datasets is therefore also unsurprising.

Interestingly, a large proportion of genes displaying evidence of allelic imbalance showed differences between the wild-type and mutant genotypes: many of the genes showing evidence of allelic imbalance in wild-type flies did not show allelic imbalance in mutant flies, and vice versa (Figure 5, Supplementary Figure 3). Nevertheless, there did not appear to be any consistent bias in the expression of one background's alleles over the other, nor did there to appear to be any change in the overall degree of allelic imbalance between wild-type and $sd^{E3}$ flies (Figure 5, Supplementary Figure 3). The latter finding is somewhat surprising, given that we might expect reduced SD levels in $sd^{E3}$ flies to minimize the impact of polymorphic SD binding sites on the expression of SD target genes. However, this analysis was also limited by small sample size; we only had polymorphic sequence tags for a handful of genes, and many of these were expressed at low levels. Larger-scale RNA-seq experiments with longer reads capable of distinguishing alleles will be able to shed more light on this matter. Combined, though, the overall picture that emerges from the binding site and allelic imbalance analyses suggests that cis-regulatory variants between SAM and ORE contribute little to the background dependence of the $sd^{E3}$ phenotype, and that variation in trans-acting factors may be more important.

**Candidate genes:** We identified several loci for further study (Table 4). This list contains a number of genes for which a role in wing phenotypes is logical based on prior evidence (e.g., *vg*, *sbb*, *bi*/*Omb*, *dlg1*, *ban*), suggesting that our approach is selecting a reasonable list of candidate genes. The presence of several genes with little or no experimental evidence for a function in wing development (e.g., *msk*, *obst-B*) shows that this strategy is also capable of detecting novel candidates. This is a key point if our goal is to develop a detailed, unbiased picture of the genetic networks underlying phenotypic variation.

## Conclusions

The genetic basis of variation in the penetrance and expressivity of mutations is likely to be complex. Dissecting the genetic basis of this background dependence will require an integrative approach, because precisely mapping alleles with small quantitative effects is difficult, even in well developed model species like *Drosophila melanogaster*. Drawing on data from multiple experiments with distinct approaches, we have identified a robust set of candidate genes for further investigation as possible factors underlying the background dependence of *scalloped*$^{E3}$'s variable phenotypic effects on the *D. melanogaster* wing. This approach should prove useful in understanding how genetic background interacts with specific mutations to influence organismal phenotypes.

**Competing Interests**

The authors declare that they have no competing interests.

**Authors' Contributions**

CHC prepared DNA for sequencing, conducted analyses, and drafted the manuscript. SC conducted the modifier deletion screen. ID & DT performed the digital gene expression experiment and assisted with analyses. ID, DT, and SC generated the backcross introgression lines. ID conceived and designed the study and participated in analyses and writing. All authors read and approved the final manuscript.

**Acknowledgements**

We thank Casey Bergman, Titus Brown, and Greg Gibson for valuable discussions on analyses and for helpful suggestions on earlier versions of this manuscript. This work was supported by NSF MCB-0922344 (to ID). This material is based in part upon work supported by the National Science Foundation under Cooperative Agreement No. DBI-0939454. Any opinions, findings, and conclusions or recommendations expressed in this material are those of the author(s) and do not necessarily reflect the views of the National Science Foundation.

**Figure 1.** Phenotypic effects of the *sd^E3* mutation on the *Drosophila melanogaster* wing in the Oregon-R (ORE) and Samarkand (SAM) genetic backgrounds at 40X magnification.  While SAM and ORE wild-type wings are qualitatively similar, the phenotypic effects of the *sd^E3* are profoundly different across these two wild-type backgrounds.

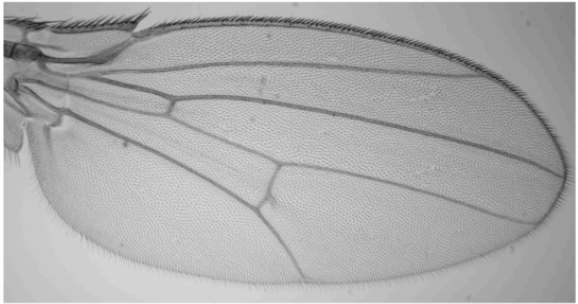

**Oregon-R wild-type**

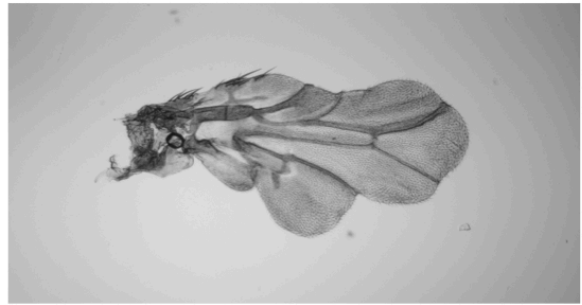

**Oregon-R *sd^E3***

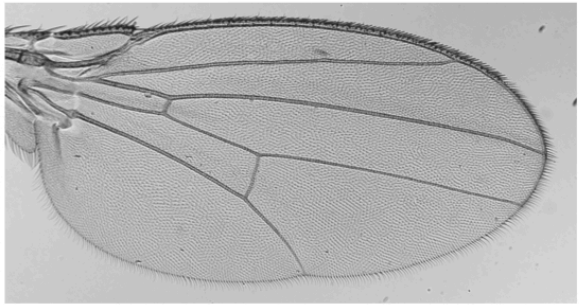

**Samarkand wild-type**

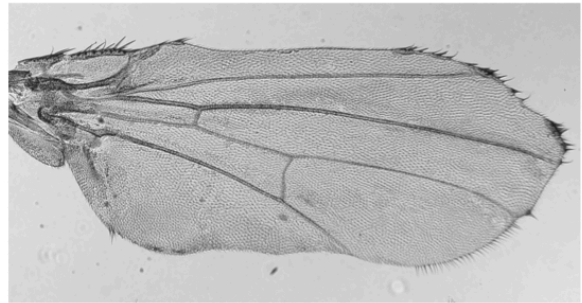

**Samarkand *sd^E3***

**Figure 2.** Backcross selection strategy used to introgress alleles from one background into another. The $sd^{E3}$ allele (insertion denoted by grey triangle) was originally backcrossed into both the Samarkand background (blue) with the relatively less severe phenotype, and Oregon-R (red) with higher expressivity (top row). The light blue genomic fragment proximal to the $sd^{E3}$ allele represents the genomic region from the original genetic background that $sd^{E3}$ was generated on (neither SAM nor ORE), while the gray coloration in generation three and beyond indicates unknown chromosomal composition due to recombination between the SAM and ORE chromosomes in previous generations. The two-generation cycle consisted of generating F2s and selecting flies with the smallest wings, and subsequently crossing these back to a pure SAM $sd^{E3}$ background. This cycle was repeated 12 times and should theoretically replace ~95% of the genome, with the exception of the regions that contribute to the background-dependent variation for expressivity of the mutant phenotype.

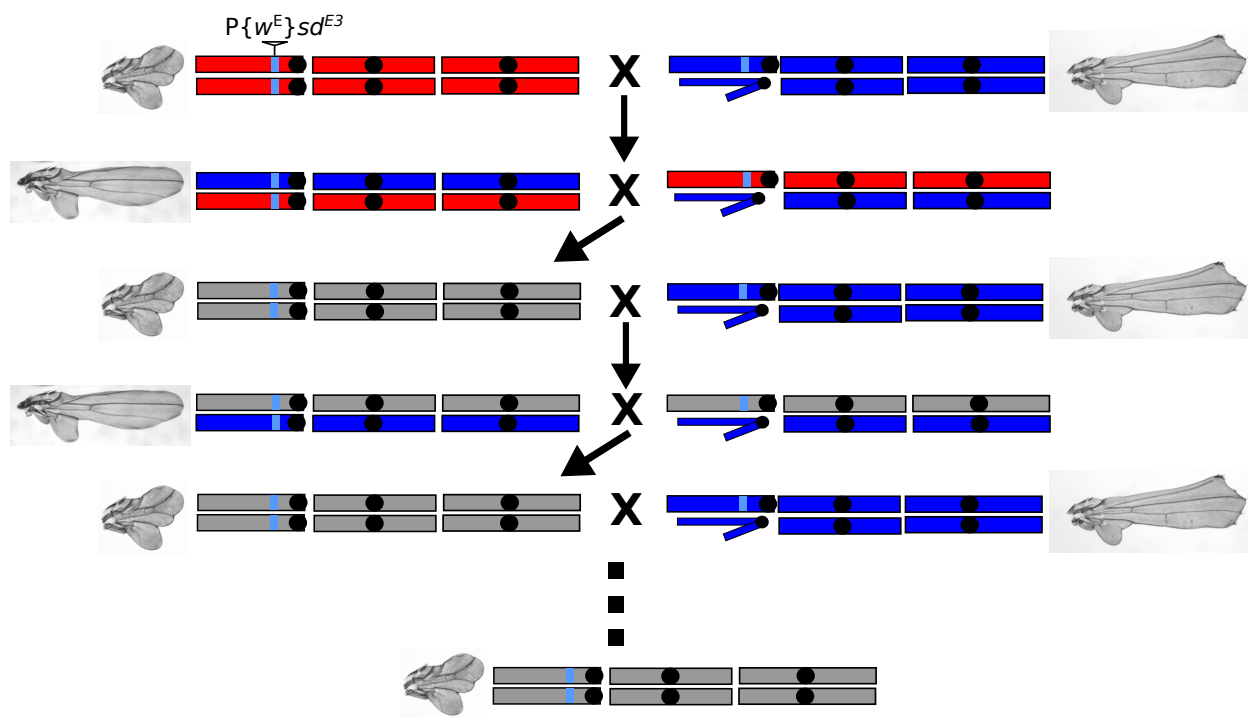

**Figure 3.** Comparison of wing phenotypes of parental (SAM and ORE) $sd^{E3}$ flies, and of long- and short-wing backcross lines generated for this study, confirming phenotypic similarity between SAM and long-wing backcross line, and between ORE and short-wing backcross lines.

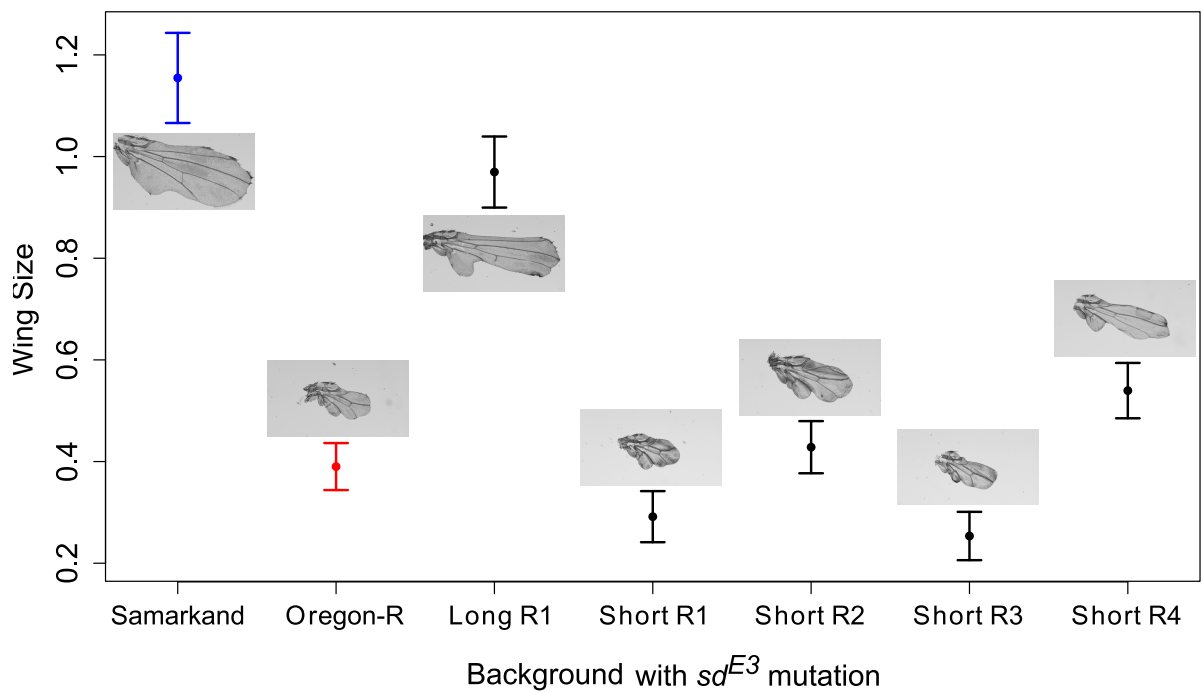

**Figure 4.** Introgression of genomic fragments among the selection-backcross lineages. (A) Idealized expectations for the frequency of the ORE allele along the length of a chromosome arm in short- and long-wing $sd^{E3}$ backcross lines; spikes in allele frequencies should indicate loci contributing to the background-dependent wing phenotype differences. (B, C) Observed frequency of the ORE allele along the length of chromosome arms X and 2L in short- and long-wing $sd^{E3}$ backcross lines. In B, the peak corresponds to the position of the introgression of the $sd^{E3}$ mutation, and thus represents a good positive control that the method is working. Allele frequency spikes are expected around $sd$ because the $sd^{E3}$ allele was originally introgressed into SAM and ORE from another genetic background, and therefore the surrounding region still carries the haplotype of the $sd^{E3}$ progenitor strain, containing a mixture of SNPs that distinguish SAM and ORE. In C, variation in the extent of the introgressed fragment (with independent breakpoints) is observed.

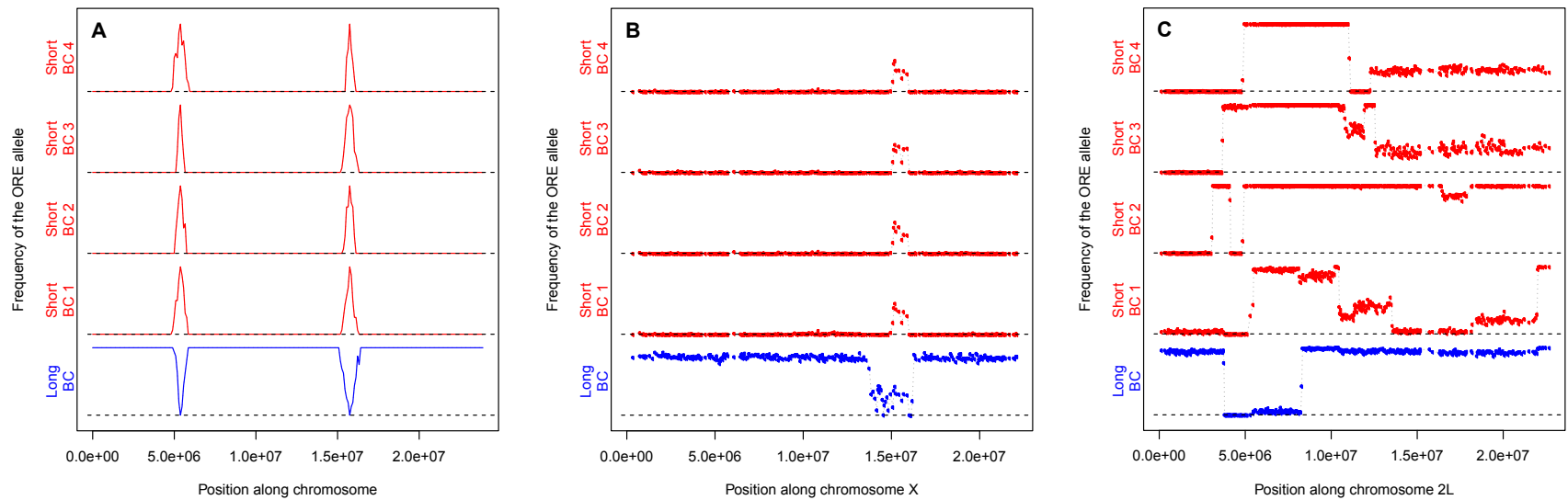

**Figure 5. Allelic imbalance does not show a general pattern of perturbation in the** $sd^{E3}$ **mutants.** Allelic imbalance in SAM/ORE "hybrid" flies with both wild-type and $sd^{E3}$ genotypes. Only sequence tags represented by at least five reads of each allele in all samples are represented in this plot.

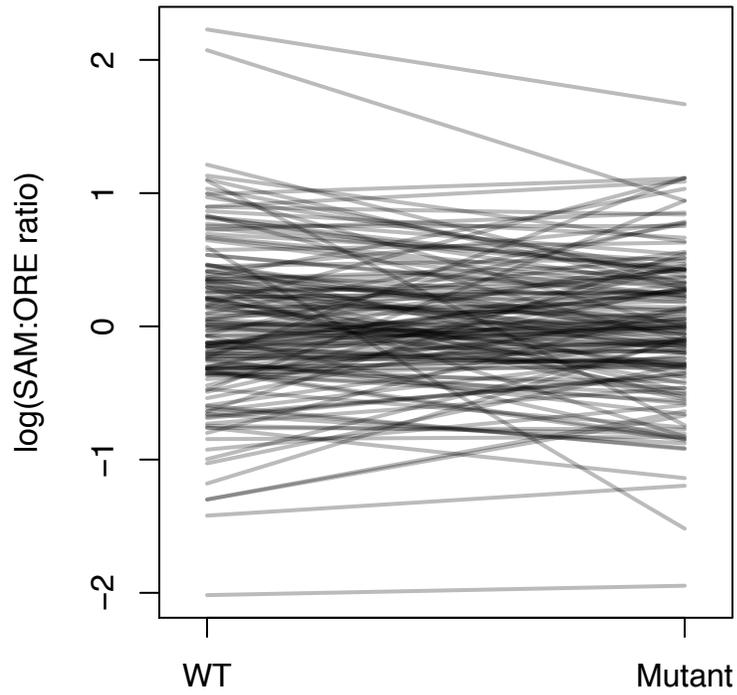

**Figure 6. Integration of multiple genomic data sets to generate a narrow set of candidate loci influencing background dependence of $sd^{E3}$.** Examples of integrated plots showing results of independent genomic datasets used to investigate the genetic basis of background dependence of the $sd^{E3}$ phenotype. Backcross: Average frequency of the ORE (short-wing) allele across four short-wing introgression lines. Modifier deletions: open bars represent deletions with a significant main effect on the $sd^{E3}$ phenotype; light shaded bars represent deletions with a significant background-dependent effect on the $sd^{E3}$ phenotype; and dark shaded bars represent deletions in which both the main and interaction effects are significant. DGE: open bars represent genes whose transcript counts are influenced by $sd$ genotype; light shaded bars represent genes showing evidence of genotype-dependent allelic imbalance; and dark shaded bars represent genes showing evidence of both an overall effect of $sd$ genotype and genotype-dependent allelic imbalance. DGRC and Illumina: open bars represent genes showing evidence of an effect of $sd$ genotype on expression; light shaded bars represent genes showing evidence of a genotype-by-background interaction effect; and dark shaded bars represent genes showing evidence of both the main and interaction effects. Binding predictions: open bars represent genes predicted to be overall SD binding targets (in at least one of the two genetic backgrounds); light shaded bars represent genes predicted to show differential affinity for SD between the two backgrounds; and dark shaded bars represent genes showing evidence of both overall SD binding and differential affinity between backgrounds. Only genes showing evidence of at least four significant effects across all datasets are shown.

**Table 1.** Top 50 genes predicted to be targets of regulation by SD in the SAM and ORE genetic backgrounds.

| Gene | p$_{SAM}$ | p$_{ORE}$ | p$_{Diff}$ |
|---|---|---|---|
| nej | 0.001 | 0.003 | 0.423 |
| CG3795 | 0.001 | 0.001 | 0.043 |
| 5SrRNA-Psi:CR33371 | 0.001 | 0.001 | 0.39 |
| Uch-L5R | 0.001 | 0.004 | 0.001 |
| cdi | 0.002 | 0.001 | 0.001 |
| 5SrRNA:CR33434 | 0.001 | 0.001 | 0.048 |
| CG14752 | 0.005 | 0.001 | 0.007 |
| Apf | 0.002 | 0.001 | 0.013 |
| Muc14A | 0.001 | 0.001 | 0.001 |
| Muc12Ea | 0.001 | 0.001 | 0.001 |
| na | 0.001 | 0.004 | 0.001 |
| Cht6 | 0.001 | 0.001 | 0.038 |
| PGRP-LB | 0.001 | 0.001 | 0.031 |
| CR43211 | 0.001 | 0.001 | 0.008 |
| CG6084 | 0.002 | 0.006 | 0.148 |
| Myo10A | 0.003 | 0.002 | 0.103 |
| 5SrRNA:CR33370 | 0.002 | 0.002 | 0.11 |
| 5SrRNA:CR33377 | 0.002 | 0.002 | 0.072 |
| CG32669 | 0.002 | 0.003 | 0.155 |
| CG11180 | 0.002 | 0.002 | 0.03 |
| feo | 0.002 | 0.002 | 0.05 |
| Hexo2 | 0.002 | 0.003 | 0.09 |
| 5SrRNA:CR33433 | 0.002 | 0.002 | 0.091 |
| 5SrRNA:CR33445 | 0.002 | 0.002 | 0.156 |
| 5SrRNA:CR33447 | 0.002 | 0.002 | 0.135 |
| Trxr-1 | 0.002 | 0.003 | 0.009 |
| rtv | 0.005 | 0.002 | 0.002 |
| ND75 | 0.002 | 0.002 | 0.006 |
| Reg-2 | 0.007 | 0.002 | 0.005 |
| CG42832 | 0.002 | 0.002 | 0.002 |
| CG31907 | 0.004 | 0.002 | 0.009 |
| CG16742 | 0.002 | 0.003 | 0.163 |
| CG17287 | 0.002 | 0.005 | 0.116 |
| Rcd6 | 0.004 | 0.002 | 0.017 |
| Rab9Fb | 0.004 | 0.002 | 0.06 |
| CG13917 | 0.004 | 0.002 | 0.032 |
| CG7530 | 0.007 | 0.003 | 0.001 |
| CG14551 | 0.003 | 0.004 | 0.001 |
| snoRNA:Me18S-G962 | 0.008 | 0.003 | 0.001 |
| Bx | 0.011 | 0.003 | 0.002 |
| Trx-2 | 0.013 | 0.003 | 0.002 |
| sofe | 0.003 | 0.004 | 0.036 |
| ush | 0.004 | 0.003 | 0.019 |
| 5SrRNA:CR33372 | 0.003 | 0.003 | 0.132 |
| 5SrRNA:CR33448 | 0.003 | 0.003 | 0.137 |
| 5SrRNA:CR33435 | 0.003 | 0.003 | 0.041 |
| 5SrRNA:CR33450 | 0.003 | 0.004 | 0.19 |
| CG17290 | 0.008 | 0.003 | 0.004 |
| CG6481 | 0.003 | 0.005 | 0.073 |
| CG13056 | 0.003 | 0.006 | 0.045 |

$P_{SAM}$, $P_{ORE}$: p-value for being a predicted SD-binding target in SAM and ORE, respectively. $P_{diff}$: p-value for differential SD binding affinity between the SAM and ORE backgrounds.

**Table 2.** Genes showing evidence of genotype-dependent allelic imbalance in SAM/ORE "hybrid" wild-type and $sd^{E3}$ mutant flies.

| Gene | ORE count (mut. 1) | SAM count (mut. 1) | ORE count (mut. 2) | SAM count (mut. 2) | ORE count (WT 1) | SAM count (WT 1) | ORE count (WT 2) | SAM count (WT 2) | p | q |
|------|------|------|------|------|------|------|------|------|------|------|
| CG6308 | 0 | 115 | 0 | 109 | 96 | 0 | 0 | 0 | 5.13E-87 | 6.50E-84 |
| Reg-2 | 92 | 2 | 69 | 5 | 33 | 36 | 52 | 34 | 7.51E-20 | 4.76E-17 |
| CG12708 | 2 | 117 | 0 | 0 | 38 | 31 | 0 | 0 | 5.17E-19 | 2.18E-16 |
| CG3632 | 1 | 41 | 0 | 55 | 26 | 31 | 30 | 28 | 2.39E-18 | 7.57E-16 |
| CG6739 | 7 | 1 | 9 | 2 | 0 | 17 | 0 | 21 | 8.79E-13 | 2.23E-10 |
| mthl10 | 8 | 0 | 9 | 0 | 0 | 5 | 0 | 6 | 9.05E-10 | 1.91E-07 |
| mtm | 9 | 0 | 5 | 0 | 0 | 7 | 0 | 5 | 2.09E-09 | 3.78E-07 |
| CG4452 | 14 | 8 | 5 | 12 | 22 | 0 | 17 | 0 | 3.72E-09 | 5.89E-07 |
| CG33785 | 2 | 5 | 5 | 7 | 19 | 0 | 15 | 0 | 1.81E-08 | 2.52E-06 |
| CG6310 | 75 | 82 | 85 | 75 | 50 | 102 | 30 | 102 | 1.99E-08 | 2.52E-06 |
| CG2147 | 13 | 0 | 19 | 0 | 26 | 0 | 0 | 25 | 2.73E-08 | 3.14E-06 |
| CG9281 | 0 | 14 | 0 | 17 | 7 | 9 | 8 | 3 | 6.49E-08 | 6.85E-06 |
| CG1908 | 16 | 3 | 16 | 4 | 4 | 27 | 11 | 12 | 9.89E-08 | 9.64E-06 |
| Spt5 | 93 | 45 | 74 | 55 | 81 | 105 | 73 | 111 | 1.66E-07 | 1.45E-05 |
| mth | 65 | 13 | 0 | 0 | 49 | 57 | 63 | 51 | 1.72E-07 | 1.45E-05 |
| CG1140 | 30 | 6 | 21 | 5 | 8 | 13 | 10 | 20 | 2.16E-07 | 1.71E-05 |
| CG1542 | 10 | 14 | 14 | 15 | 33 | 3 | 22 | 4 | 3.03E-07 | 2.26E-05 |
| CG6051 | 15 | 2 | 8 | 3 | 5 | 25 | 11 | 17 | 1.03E-06 | 6.89E-05 |
| U3-55K | 0 | 0 | 32 | 62 | 42 | 22 | 34 | 14 | 1.03E-06 | 6.89E-05 |
| CG5704 | 4 | 0 | 2 | 0 | 0 | 7 | 0 | 6 | 1.13E-06 | 7.14E-05 |
| CG32202 | 1 | 8 | 0 | 9 | 12 | 1 | 11 | 9 | 2.31E-06 | 0.000139294 |
| CG5626 | 1 | 19 | 1 | 18 | 9 | 12 | 11 | 8 | 2.42E-06 | 0.000139294 |
| CG16888 | 0 | 2 | 0 | 3 | 7 | 0 | 7 | 0 | 2.87E-06 | 0.000158171 |
| scaf6 | 0 | 3 | 0 | 3 | 3 | 0 | 7 | 0 | 4.20E-06 | 0.000221877 |
| CG16799 | 0 | 18 | 0 | 12 | 7 | 8 | 5 | 9 | 6.75E-06 | 0.000342299 |
| RpL7-like | 0 | 0 | 345 | 239 | 0 | 0 | 279 | 325 | 8.50E-06 | 0.000414227 |
| CAH2 | 4 | 10 | 0 | 0 | 13 | 0 | 8 | 1 | 1.04E-05 | 0.000490105 |

| | | | | | | | | | |
|---|---|---|---|---|---|---|---|---|---|
| CG4908 | 21 | 6 | 15 | 10 | 5 | 19 | 6 | 13 | 1.62E-05 | 0.000732638 |
| CG1673 | 0 | 2 | 0 | 5 | 2 | 0 | 4 | 0 | 2.27E-05 | 0.000993509 |
| CG2260 | 0 | 0 | 120 | 34 | 122 | 69 | 106 | 88 | 2.63E-05 | 0.001088493 |
| Chc | 4 | 0 | 7 | 0 | 10 | 13 | 2 | 8 | 2.66E-05 | 0.001088493 |
| RpL7-like | 5 | 1 | 5 | 3 | 0 | 7 | 2 | 16 | 2.94E-05 | 0.001164226 |
| CG30286 | 0 | 6 | 0 | 5 | 2 | 0 | 2 | 0 | 3.03E-05 | 0.001164226 |
| CG9005 | 0 | 3 | 0 | 5 | 0 | 0 | 5 | 0 | 3.15E-05 | 0.001174983 |
| Pep | 1 | 3 | 3 | 8 | 10 | 0 | 0 | 0 | 3.94E-05 | 0.001427063 |
| CG12424 | 3 | 0 | 5 | 1 | 2 | 22 | 5 | 10 | 4.53E-05 | 0.001510213 |
| Khc-73 | 2 | 0 | 4 | 0 | 0 | 3 | 0 | 3 | 4.53E-05 | 0.001510213 |
| CG13749 | 0 | 3 | 0 | 3 | 4 | 0 | 2 | 0 | 4.53E-05 | 0.001510213 |
| CG12009 | 2 | 2 | 2 | 4 | 8 | 0 | 12 | 0 | 4.70E-05 | 0.00152812 |
| D12 | 15 | 5 | 0 | 0 | 5 | 19 | 5 | 16 | 5.00E-05 | 0.001558229 |
| Aats-glupro | 8 | 1 | 4 | 0 | 1 | 3 | 1 | 7 | 5.04E-05 | 0.001558229 |
| CG5144 | 3 | 0 | 2 | 0 | 0 | 2 | 0 | 5 | 5.40E-05 | 0.00161654 |
| Lap1 | 3 | 4 | 10 | 4 | 34 | 2 | 30 | 0 | 5.59E-05 | 0.00161654 |
| CG17841 | 24 | 4 | 17 | 3 | 15 | 19 | 21 | 16 | 5.61E-05 | 0.00161654 |
| Eip71CD | 0 | 0 | 9 | 0 | 0 | 0 | 0 | 4 | 6.17E-05 | 0.001697853 |
| CG7376 | 0 | 0 | 0 | 4 | 5 | 0 | 4 | 0 | 6.17E-05 | 0.001697853 |
| CG10650 | 4 | 1 | 4 | 0 | 0 | 5 | 0 | 2 | 6.67E-05 | 0.001697853 |
| CG15382 | 1 | 1 | 6 | 0 | 0 | 5 | 0 | 3 | 6.67E-05 | 0.001697853 |
| Not1 | 4 | 1 | 3 | 0 | 0 | 6 | 0 | 2 | 6.67E-05 | 0.001697853 |
| CG5189 | 6 | 1 | 6 | 0 | 0 | 1 | 0 | 4 | 6.81E-05 | 0.001697853 |

P-values are derived from likelihood ratio tests comparing a model with genotype-dependent allelic imbalance to a model with the same level of allelic imbalance in wild-type and $sd^{E3}$ flies, and q-values were computer using the qvalue package in R. Top fifty genes are shown.

**Table 3.** Results of tests to distinguish between second- and higher-order epistasis to account for background-dependent dominant modifiers of the $sd^{E3}$ phenotype.

| Deletion | Deletion name | Chrom | Start | End | S1.SAM | S1.ORE | S3.SAM | S3.ORE | L1.SAM | L1.ORE |
|---|---|---|---|---|---|---|---|---|---|---|
| 6326 | Control | NA | NA | NA | -2.75 to -2.39, ORE | -0.11 to 0.21, ORE | -2.69 to -2.3, ORE | -0.21 to 0.16, ORE | -0.47 to 0.07, SAM | 2.06 to 2.58, SAM |
| 7491 | Df(2L)Exel6004 | 2L | 1074079 | 1158137 | -1.81 to -0.37, SAM* | -0.82 to 0, SAM* | -2.23 to -0.34, SAM* | -0.79 to 0.36, ORE | -1.48 to 0.63, ORE* | 0.31 to 1.84, ORE* |
| 7492 | Df(2L)Exel6005 | 2L | 1555098 | 1737249 | -3.14 to -2.57, SAM* | 0.69 to 2.36, SAM | -3.65 to -2.92, SAM* | 1.08 to 2.82, ORE | -0.43 to 0.14, ORE* | 0.36 to 2.03, ORE* |
| 7503 | Df(2L)Exel6017 | 2L | 7202317 | 7418128 | -1.72 to -0.95, ORE | 0.63 to 1.14, ORE* | -1.49 to -0.32, ORE | 0.03 to 0.89, ORE | -0.49 to 0.68, SAM | 0.11 to 0.97, SAM |
| 7505 | Df(2L)Exel6021 | 2L | 8989308 | 9176164 | -1.6 to -0.33, Ambig | -0.35 to 1.11, Ambig | -1.12 to -0.18, ORE | -0.47 to 0.6, ORE | -0.56 to 0.39, ORE* | -0.04 to 1.04, ORE |
| 7516 | Df(2L)Exel6033 | 2L | 12423459 | 12655793 | -1.03 to -0.04, Ambig | -0.79 to 0.36, Ambig | -1.87 to -0.3, Ambig | -0.59 to 1.25, Ambig | -0.44 to 0.61, ORE* | 0.22 to 1.45, ORE* |
| 7524 | Df(2L)Exel6042 | 2L | 18973942 | 19161727 | -3.14 to -1.14, Ambig | -1.38 to 0.81, Ambig | -2.69 to -1.06, Ambig | -1.49 to 0.38, Ambig | NA | 1.83 to 3.02, ORE* |
| 7525 | Df(2L)Exel6043 | 2L | 19161727 | 19423709 | -2.58 to -0.97, Ambig | -1.3 to 0.85, Ambig | -3.02 to -1.26, Ambig | -1.06 to 1.35, Ambig | -0.91 to 0.34, ORE* | 0.68 to 2.74, ORE* |
| 7529 | Df(2L)Exel6047 | 2L | 21102742 | 21244119 | -2.67 to -1.38, Ambig | -0.65 to 1.7, Ambig | -2.97 to -1.14, Ambig | -0.69 to 1.8, Ambig | -0.47 to 0.14, ORE* | 0.31 to 2.36, ORE* |
| 7540 | Df(2R)Exel6058 | 2R | 4215033 | 4332249 | -2.61 to -1.71, ORE | -1.26 to 0.2, ORE | -2.77 to -1.84, ORE | -1.13 to 0.35, ORE | -0.81 to 0.06, Ambig | 1.48 to 3.15, Ambig |
| 7541 | Df(2R)Exel6059 | 2R | 6761890 | 7073552 | -1.22 to -0.27, Ambig | -0.14 to 0.5, Ambig | -0.99 to 0.07, Ambig | -0.43 to 0.22, Ambig | -0.27 to 0.63, Ambig | 0.41 to 1.08, Ambig |
| 7544 | Df(2R)Exel6062 | 2R | 8868689 | 8922684 | -2.71 to -1.58, Ambig | -1 to 0.74, Ambig | -2.47 to -1.84, Ambig | -0.72 to 0.48, Ambig | -1.34 to -0.09, SAM* | 0.67 to 2.45, SAM |
| 7554 | Df(2R)Exel6072 | 2R | 16944303 | 17138350 | -2.95 to -2.05, SAM* | -1.07 to 0.57, SAM* | -3.05 to -2.2, SAM* | -0.85 to 0.6, SAM* | -1.5 to 0.36, SAM | 1.13 to 3.22, SAM |
| 7585 | Df(3L)Exel6106 | 3L | 5601375 | 5684102 | -1.18 to -0.38, ORE | NA | -1.14 to -0.41, SAM* | 0 to 0, SAM* | -0.56 to 0.33, ORE* | 0.29 to 1.05, ORE* |
| 7611 | Df(3L)Exel6132 | 3L | 17414682 | 17526127 | -2.24 to -0.74, ORE | -0.17 to 0.95, ORE | -2.24 to -0.74, ORE | -0.17 to 0.95, ORE | -1.08 to 0.91, Ambig | 0.17 to 1.86, Ambig |
| 7625 | Df(3R)Exel6146 | 3R | 2988409 | 3317319 | -2.36 to 0.67, Ambig | -1.59 to 1.73, Ambig | -2.85 to -0.69, Ambig | -0.4 to 2.4, Ambig | -0.43 to 2.03, Ambig | -0.19 to 3.33, Ambig |
| 7677 | Df(3R)Exel6198 | 3R | 19967091 | 20096927 | -2.07 to -0.96, Ambig | -1.18 to -0.06, Ambig | -3.06 to -1.88, SAM* | -0.26 to 0.93, SAM* | -0.13 to 0.53, ORE* | 1.95 to 2.72, ORE* |
| 7686 | Df(3R)Exel6208 | 3R | 22983208 | 23079244 | -0.78 to 0.05, SAM | -0.28 to 0.34, SAM* | -0.49 to 0.04, SAM | -0.32 to 0.08, SAM* | -0.47 to 0.03, SAM | -0.08 to 0.31, SAM* |
| 7724 | Df(2L)Exel6256 | 2L | 5555049 | 5658629 | -3.06 to -2.29, ORE | -0.34 to 0.39, ORE | -3.3 to -2.34, ORE | -0.3 to 0.63, ORE | -0.19 to 0.84, SAM | 2.49 to 3.47, SAM |
| 7741 | Df(3R)Exel6274 | 3R | 19001169 | 19121356 | -2.33 to -0.67, SAM* | -1.2 to 0.78, SAM* | -2.91 to -1.66, SAM* | -0.45 to 1.59, SAM* | -0.45 to 0.17, ORE* | 0.68 to 2.46, ORE* |
| 7795 | Df(2L)Exel7021 | 2L | 4915628 | 4979299 | -3.26 to -0.94, SAM* | -0.14 to 0.47, SAM* | -3.3 to -1.3, ORE | -0.06 to 0.8, ORE | -1.28 to 2.41, SAM | 1.85 to 3.15, SAM |
| 7834 | Df(2L)Exel7067 | 2L | 16728375 | 16824908 | -2.86 to -1.71, SAM* | -1.14 to 1.99, SAM* | -3.51 to -1.97, Ambig | -0.49 to 2.26, Ambig | -2.41 to -0.16, ORE | -1.19 to 2.33, ORE |

| 7845 | Df(2L)Exel7073 | 2L | 18859186 | 19022139 | -2.63 to -1.66, Ambig | -1.1 to 0.31, Ambig | -2.93 to -2.21, Ambig | -0.55 to 0.62, Ambig | -0.26 to 0.26, ORE* | 2.03 to 3.04, ORE* |
|---|---|---|---|---|---|---|---|---|---|---|
| 7875 | Df(2R)Exel7130 | 2R | 9960585 | 10100288 | -2.47 to -1.69, Ambig | -0.91 to 0.66, Ambig | -3.13 to -2.28, Ambig | -0.36 to 1.37, Ambig | 0.03 to 0.65, SAM* | 1.97 to 3.12, SAM |
| 7879 | Df(2R)Exel7135 | 2R | 11017461 | 11150447 | -1.49 to -0.22, ORE | NA | -1.6 to -0.12, Ambig | 0 to 0, Ambig | -1.06 to 0.47, SAM | -0.06 to 1.19, SAM* |
| 7894 | Df(2L)BSC50 | 2R | 14509027 | 14618276 | -3.25 to -2.51, ORE | 0.08 to 1.07, SAM* | -3.38 to -2.39, SAM* | -0.09 to 1.23, ORE* | -0.39 to 0.38, ORE* | 1.81 to 2.8, ORE* |
| 7949 | Df(3L)Exel9065 | 3L | 21510121 | 21516264 | -0.76 to -0.04, ORE | -0.35 to 0.35, ORE | -1.19 to -0.22, ORE | -0.11 to 0.72, ORE* | -0.37 to 0.14, ORE* | -0.07 to 0.64, ORE* |
| 7960 | Df(3R)Exel7309 | 3R | 7472850 | 7541866 | -1.27 to 1.56, Ambig | -1.98 to 1.98, Ambig | -1.81 to -0.48, ORE | -0.39 to 2.97, ORE | -0.49 to 2.15, Ambig | -1.35 to 2.73, Ambig |
| 7976 | Df(3R)Exel8159 | 3R | 9809236 | 10085649 | -1.06 to -0.5, Ambig | NA | -1.04 to -0.51, ORE | 0 to 0, ORE | -0.97 to 0.27, ORE | -0.22 to 1.08, ORE |
| 7993 | Df(3R)Exel8178 | 3R | 20096927 | 20353553 | -2.83 to -1.43, Ambig | 0.12 to 1.33, SAM* | -3.85 to -2.05, Ambig | 0.7 to 2.4, SAM | -2.48 to -0.25, ORE | -0.92 to 0.99, ORE |
| 7995 | Df(3R)Exel9025 | 3R | 25570975 | 25585907 | -3.01 to -0.56, SAM* | 0 to 2.23, SAM* | -0.96 to -0.38, SAM* | 0 to 0, SAM* | -0.82 to 0.35, ORE* | 0.02 to 0.83, ORE |
| 7998 | Df(2L)BSC50 | 2R | 16758362 | 16887468 | -3.08 to -2.42, SAM* | -0.4 to 0.57, SAM* | -4.15 to -2.3, SAM* | -0.41 to 1.52, SAM* | -3.39 to -1.71, SAM* | -0.85 to 1.08, SAM* |

Short- and long-backcross strains were crossed to deletion mutants, and wing phenotypes were compared to those of offspring from crosses between the deletion and the parental SAM $sd^{E3}$ or ORE $sd^{E3}$ strains. Deletion, Deletion name, Chrom, Start, and End provide the identities and chromosomal locations of the deletions tested. Remaining columns indicate results of pairwise comparisons between backcross × deletion flies (S1 = short backcross 1, S3 = short backcross 3, L1 = long backcross) and the parental × deletion flies (SAM or ORE). Numbers indicate 95% confidence intervals for mean differences in wing phenotypes in each comparison, and "ORE", "SAM", and "Ambig" inside each cell indicates the hemizygous allele carried by that backcross strain at the deletion locus. Asterisks indicate comparisons in which the results rule out second-order epistasis (i.e., flies have indistinguishable wing phenotypes but different alleles opposite the deletion, or different phenotypes but the same allele).

**Table 4.** Candidate genes identified by integrative analysis of gene expression, sequence binding predictions, and mapping and modifier datasets.

| Gene name | Chrom. | Pos. | Evidence |
|---|---|---|---|
| Prosap | 2R | 9947961 .. 10028407 | Expression, introgression, modifier deletions, binding predictions |
| CG9427 | 3R | 5450879 .. 5452652 | Expression, introgression, modifier deletions |
| eca | 3R | 5510191 .. 5511198 | Expression, introgression, modifier deletions, binding predictions |
| ihog | 2L | 6945464 .. 6948808 | Expression, introgression, modifier deletions |
| obst-B | 2L | 10032690 .. 10038954 | Expression, introgression, modifier deletions |
| lola | 2R | 6369399 .. 6430796 | Expression, introgression, modifier deletions, binding predictions |
| sbb | 2R | 14165703 .. 14244801 | Expression, introgression, modifier deletions |
| vg | 2R | 8772137 .. 8786890 | Expression, introgression, modifier deletions |
| Vps45 | 3R | 5389646 .. 5391814 | Expression, introgression, modifier deletions, binding predictions |
| pont | 3R | 6087845 .. 6089621 | Expression, introgression, modifier deletions |
| CG17230 | 3R | 7263507 .. 7293094 | Expression, introgression, modifier deletions, binding predictions |